\documentclass{aa}
\usepackage{amsmath}
\usepackage{amssymb}
\usepackage{graphicx}
\usepackage{txfonts}
\usepackage{hyperref}  % To add links in the article.
\usepackage{multirow}
\usepackage{float}
\usepackage{orcidlink}
\usepackage{soul}
\usepackage{xcolor}
\hypersetup{colorlinks = true, allcolors = blue}
\usepackage{CJKutf8}
\setul{0.3ex}{1pt} % 设置红色下划线属性：与文本距离，粗细

\begin{document}
    \title{Search for quasar pairs with \textit{Gaia} astrometric data.}
    \subtitle{III. Discovery of 9 dual and projected quasars}
    \titlerunning{Search for Quasar Pairs with \textit{Gaia} Astrometric Data. III.}
   \author{
           Qihang Chen (\begin{CJK}{UTF8}{gbsn}{陈启航}\end{CJK}) \inst{\ref{BNU_PA},\ref{BNU_FIAA}} \orcidlink{0009-0006-9345-9639}
           % \email{202131160006@mail.bnu.edu.cn}
           \and
           Zizhao He (\begin{CJK}{UTF8}{gbsn}{何紫朝}\end{CJK}) \inst{\ref{NCU_DOP},\ref{NCU_CRAHEP},\ref{PMO}} \orcidlink{0000-0001-8554-9163}
           % \email{zzhe@ncu.edu.cn}
           \and
           Zhuojun Deng (\begin{CJK}{UTF8}{gbsn}{邓卓君}\end{CJK}) \inst{\ref{BNU_PA},\ref{BNU_FIAA}} \orcidlink{0009-0008-8080-3124}
           % \email{202431101076@mail.bnu.edu.cn}
           \and
           Liang Jing (\begin{CJK}{UTF8}{gbsn}{荆亮}\end{CJK}) \inst{\ref{BNU_PA},\ref{BNU_FIAA}} \orcidlink{0000-0003-1188-9573}
           % \email{202331160009@mail.bnu.edu.cn}
           \and \\
           Xingyu Zhu (\begin{CJK}{UTF8}{gbsn}{朱星宇}\end{CJK}) \inst{\ref{BNU_PA},\ref{BNU_FIAA}} \orcidlink{0009-0008-9072-4024}
           % \email{202321160028@mail.bnu.edu.cn}
           \and
           Jianghua Wu (\begin{CJK}{UTF8}{gbsn}{吴江华}\end{CJK}) \inst{\ref{BNU_PA},\ref{BNU_FIAA}} \thanks{Corresponding authors: \href{mailto:jhwu@bnu.edu.cn}{jhwu@bnu.edu.cn}} \orcidlink{0000-0002-8709-6759}
           }
   \institute{
              School of Physics and Astronomy, Beijing Normal University, Beijing, 100875, China \\ \email{jhwu@bnu.edu.cn} \label{BNU_PA}
              \and
              Institute for Frontier in Astronomy and Astrophysics, Beijing Normal University, Beijing, 102206, China \label{BNU_FIAA}
              \and
              Department of Physics, Nanchang University, Nanchang, 330031, China \label{NCU_DOP}
              \and
              Center for Relativistic Astrophysics and High Energy Physics, Nanchang University, Nanchang, 330031, China \label{NCU_CRAHEP}
              \and
              Purple Mountain Observatory, Chinese Academy of Sciences, Nanjing, Jiangsu, 210023, China \label{PMO}
             }
   \date{Received XXXX, XXXX; accepted XXXX, XXXX}

% \abstract{}{}{}{}{} 
% 5 {} token are mandatory
  \abstract{We report the low-resolution long-slit spectroscopic observations and confirmations of 11 quasar pair candidates, which are selected from the MGQPC catalog presented in the first paper of our series work (hereafter, \textsc{Paper-I}) and the early version of this catalog. The spectroscopic follow-up was carried out with 5 spectrographs equipped on 3 telescopes, and the major discoveries include 6 dual quasars and 3 projected quasars. One of the dual quasars has a high redshift of $\sim$ 3.1. The LQ hypothesis of 3 dual quasars cannot be completely ruled out. We investigated the reason why previous spectroscopic surveys missed several new quasars. We discussed a projected quasar with a wide-separation lensing configuration, as well as two quasar-star projections that mimic the configuration of lensed quasars. The photometric redshifts for the 11 observed candidates were extracted from the second paper of our series work (hereafter, \textsc{Paper-II}) to illustrate their positive role in mitigating contamination from projected quasars and quasar-star projections. We also reviewed and discussed the confirmation strategies for dual and lensed quasar candidates, and outlined future confirmation strategies for them in the context of the era dominated by large-scale spectroscopic and imaging surveys.}
  \maketitle
  \nolinenumbers
%-------------------------------------------------------------------

\section{Introduction} \label{sec1}
% What is a quasar pair
Quasar pairs (QPs) are now widely regarded as a natural consequence of galaxy mergers and serve as key probes for studying the growth of supermassive black holes (SMBHs), galaxy formation and mass assembly, as well as the coevolution between the two host galaxies \citep[e.g.,][]{Kauffmann2000GalQSOevolution, Hopkins2005QuasarEvolution, Ellison2011SDSSGalaxyPair-IV, Ellison2015SDSSGalaxyPair-XII, Ellison2025GalaxyPost-merger-III, Volonteri2022DAGNHorizon-AGNsimulation, Pierce2023QuasarTriggering}. In the physical scenario of galaxy mergers, gravitational interactions between two massive galaxies drive gas inflows toward their respective centers, thereby triggering or enhancing SMBH accretion activity. This process typically manifests as the formation of either one or two active galactic nuclei (AGNs) at different merger stages, resulting in AGN/quasar–galaxy pairs or AGN/quasar pairs \citep[e.g.,][]{Hopkins2006DQModel, Hopkins2008DQModel, VanWassenhove2012DAGNsimulation, ChenNY2023DAGNsimulation, LiaoSH2023ModellingBSMBH, LiaoSH2024RABBITS-I, LiaoSH2024RABBITS-II, DuanQ2026GalaxyMergerAGN}. Based on the transverse distance ($r_p$) between the two quasars, QPs can be further divided into two subclasses \citep{DeRosa2019QPreview, Pfeifle2025TheBigMACDR1}: 1) dual quasars (DQs), where the $r_p$ between the members is approximately 0.03–-110 kpc, marking the stage when galaxy mergers have progressed to the intermediate or late stage and both SMBHs are significantly activated; and 2) binary quasars (BQs), with $r_p$ below 30 pc, corresponding to the final dynamical evolutionary stage before the merger of binary SMBHs, serving as unique laboratories for probing extreme gravitational environments and nanohertz gravitational waves (nHz-GWs) in the early Universe \citep[e.g.,][]{ChenYF2020BSMBHnHzGW, ShenY2023DSMBHnHzGW, ChenNY2025DAGNnHzGW}.

% QP spectroscopic follow-up, confirmation, and current number
Despite their valuable scientific significance, the observational identification of QPs remains extremely challenging. In recent years, there have been numerous efforts on the spectroscopic follow-up and confirmation of QPs. These include dedicated follow-up observations \citep[e.g.,][]{Silverman2020HSCSSP-DSMBH, TangSL2021HSCSSP-DQ, TangSL2026HSCSSP-6DQ, ZhangYW2021DAGN-II, ZhangYW2021DAGN-III, ShenY2021HSTDQ, YueMH2023LQ+QP, ChenYC2022VODKA-HSTspecObs, ChenYC2025VODKA-Gemini-HSTspecObs, Gross2025VODKA-HST-VLAspecObs, WangQ2026HST5kpcDQ, Scialpi2026CosmicDuetsDQLQ}, byproducts from the identification of lensed quasars \citep[LQs, e.g.,][]{Lemon2020STRIDES10LQ+10QPJ0229, Lemon2023GLQG150LeQ+QP+PQPJ0130J0249, HeZZ2025LeQDQPQ19}, cross-matching with large-scale spectroscopic survey databases \citep[e.g.,][]{LiuX2011LowZAGNPair, ShenY2023SDSSDR16QXGaiaEDR3-QPstatistics, JingL2025DESIDR1QP, JiangYZ2026SDSSDR16QXGaiaDR3-QPstatistics}, and self-cross-matching of known quasar catalogs \citep[]{DengZJ2026LowZAGNPair}. As a result, QPs have been discovered in large numbers, with the total number significantly exceeding that reported by \citet[][The Big MAC DR1]{Pfeifle2025TheBigMACDR1}. Based on our recent preliminary statistics, there are currently more than 1200 QPs \citep[e.g.,][]{JingL2025DESIDR1QP}, and only a few BQs have been identified \citep[e.g.,][]{Rodriguez2006BSMBH0402+379, ZhengZY2016BSMBHJ0159+0105}. The Big MAC DR2 will likely provide a more detailed catalog of confirmed QPs in the near future. Although this sample size is sufficient for preliminary statistical studies \citep[e.g.,][]{LiuX2012AGNPairStatistics, HouMC2020AGNPairXray, DeRosa2023DAGNXray, ShenY2023SDSSDR16QXGaiaEDR3-QPstatistics, JiangYZ2026SDSSDR16QXGaiaDR3-QPstatistics}, the distribution of relevant statistical parameters (such as redshift and $r_p$) remains incomplete. Such incompleteness emphasizes the urgent need for extensive spectroscopic follow-up observations for QP candidates.

% Introduce papers I and II
Motivated by the urgent need to expand the sample of spectroscopically confirmed QPs, in \textsc{Paper-I} \citep{ChenQH2025paperI} we isolated a catalog of 4\,112 candidate QPs (the MGQPC catalog) for further spectroscopic follow-up, each of which consists of members A and B. Member A is the known quasar from version 8.0 of the Million Quasar Catalog, while member B is the selected quasar candidate. Although member B in the MGQPC catalog is reported to have a quasar purity of only $\sim$ 69\,\%, the accuracy of this purity can only be verified through extensive spectroscopic follow-up observations. To preliminarily remove the contamination of projected quasars (PQs) and improve the efficiency of follow-ups of QP candidates, in \textsc{Paper-II} \citep{ZhuXY2026paperII}, we estimated photometric redshifts for a subset of quasar candidates in the MGQPC catalog. This work (\textsc{Paper-III}) reports the spectroscopic follow-up and discovery of 6 DQs and 3 PQs from the MGQPC catalog. The paper is organized as follows. The target selection, spectroscopic observations, and data reduction are described in Section \ref{sec2}. The confirmation of new DQs and PQs is described in Section \ref{sec3}. In Section \ref{sec4}, we discussed three main topics that can help optimize the strategy for DQ confirmation, aiming to provide further guidance for future spectroscopic follow-up. The summary is given in Section \ref{sec5}. Throughout this paper, we adopt a flat lambda cold dark matter ($\Lambda$CDM) cosmology with $\Omega_{\Lambda}$ = 0.7, $\Omega_M$ = 0.3, and $H_0$ = 70 km $\cdot$ s$^{-1}$ $\cdot$ Mpc$^{-1}$.
%--------------------------------------------------------------------

\section{Spectroscopic follow-up} \label{sec2}
\subsection{Target selection} \label{sec2.1}
\begin{figure*}
    \centering
    \includegraphics[width=\textwidth]{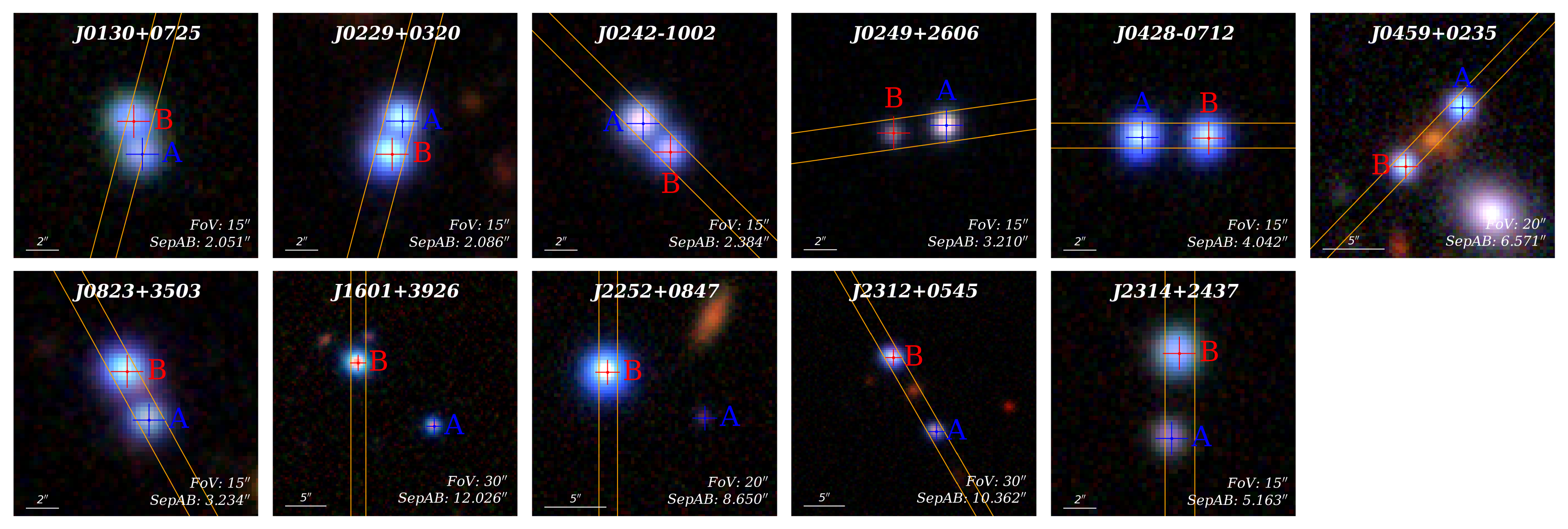}
    \caption{Pseudo-color images of the 11 observed pairs. Except for the image of J0249+2606, which comes from a Pan-STARRS composite, the images of the remaining pairs are all from DESI-LS DR10 composites. North is up and east is to the left. The slit position during the exposure is overlaid (orange double lines) in each cutout image. For all pairs except J1601+3926, J2252+0847, and J2314+2437, the spectroscopic slit was rotated, with the PA listed in Table \ref{target_obs}. The slit widths are given as used in the actual observations. ``FoV'' and ``SepAB'' marked in each pseudo-color image denote the field of view and the angular separation between the two members, respectively.}
    \label{MGQPC_Obs}
\end{figure*}

The major targets to be observed were selected from the MGQPC catalog proposed in the \textsc{Paper-I}. In addition, several targets were selected from the early version of the MGQPC catalog that was not published in \textsc{Paper-I}. Based on the seeing at the observatory sites, member angular separation, \textit{Gaia} DR3 $G$-band magnitude, and redshift of the companion known quasar, dozens of QP candidates were selected for spectroscopic follow-up. Finally, the spectra of 11 candidate QPs were obtained. Within the 11 observed pairs, 7 are selected from the MGQPC catalog. The other 4 are selected from the early version of it. Figure \ref{MGQPC_Obs} displays the pseudo-color cutout images of the spectroscopically observed pairs. All the pseudo-color images are composited from $g$-, $r$-, and $z$-band imaging data of DESI-LS DR10 or Pan-STARRS using \texttt{HumVI} \citep{Marshall2015SWHumVI, Marshall2016HumVI} via RGB combination.

\subsection{Observations} \label{sec2.2}
Low-resolution long-slit spectroscopic follow-ups for the selected targets were carried out employing 3 telescopes: Xinglong 2.16m \citep[hereafter, XLT216,][]{FanZ2016XL216OMR-BFOSC} at Xinglong Observatory of National Astronomical Observatories, Chinese Academy of Science; Lijiang 2.4m \citep[hereafter, LJT240,][]{WangCJ2019LJ240YFOSC} at Lijiang Observatory, Yunnan Observatories, Chinese Academy
of Sciences; and Palomar 200-inch (5.1 m) Hale telescope (P200\footnote{\url{https://sites.astro.caltech.edu/palomar/about/telescopes/hale.html}}) at Palomar Observatory, California. During the exposure of each pair (except two targets), the slit was rotated to a position angle (PA) to include both members simultaneously (overlaid with orange double lines in each pseudo-color cutout image in Figure \ref{MGQPC_Obs}). The detailed observation log is listed in Table \ref{target_obs}.

\begin{table*}
    \caption{Spectroscopic observation details of the 11 pairs.}
    \centering
    \begin{tabular}{c|c|c|c|c|c|c|c|c}
    \hline \hline
    Name$^{\,(1)}$ & R.A.$^{\,(2)}$ & Dec$^{\,(3)}$ & $G$\,mag$^{\,(4)}$ & SepAB$^{\,(5)}$ & Obs. Date$^{\,(6)}$ & Spectrograph$^{\,(7)}$ & Exp. Time$^{\,(8)}$ & PA$^{\,(9)}$\\ & (degree) & (degree) & (mag) & ({\arcsec}) & (yyyy-mm-dd) &  & (s) & (degree)\\
    \hline
    J0130+0725A & 22.58586 & 7.42091 & 19.514 & \multirow{2}{*}{2.051} & \multirow{2}{*}{2023-10-16} & \multirow{2}{*}{DBSP} & \multirow{2}{*}{1500} & \multirow{2}{*}{165}\\
    J0130+0725B & 22.58601 & 7.42146 & 19.018 &  &  &  &  & \\
    \hline
    J0229+0320A & 37.49238 & 3.34225 & 18.747 & \multirow{2}{*}{2.086} & \multirow{2}{*}{2022-11-15} & \multirow{2}{*}{YFOSC} & \multirow{2}{*}{2400} & \multirow{2}{*}{165}\\
    J0229+0320B & 37.49256 & 3.34170 & 18.169 &  &  &  &  & \\
    \hline
    J0242$-$1002A & 40.68886 & -10.04898 & 18.726 & \multirow{2}{*}{2.384} & \multirow{2}{*}{2023-10-16} & \multirow{2}{*}{DBSP} & \multirow{2}{*}{1800} & \multirow{2}{*}{45}\\
    J0242$-$1002B & 40.68840 & -10.04946 & 19.461 &  &  &  &  & \\
    \hline
    J0249+2606A & 42.45929 & 26.11411 & 18.802 & \multirow{2}{*}{3.210} & \multirow{2}{*}{2024-01-10} & \multirow{2}{*}{YFOSC} & \multirow{2}{*}{3600} & \multirow{2}{*}{98}\\
    J0249+2606B & 42.46027 & 26.11399 & 20.147 &  &  &  &  & \\
    \hline
    J0428$-$0712A & 67.06002 & -7.20096 & 18.675 & \multirow{2}{*}{4.042} & \multirow{2}{*}{2024-01-10} & \multirow{2}{*}{YFOSC} & \multirow{2}{*}{3600} & \multirow{2}{*}{90}\\
    J0428$-$0712B & 67.05889 & -7.20097 & 19.124 &  &  &  &  & \\
    \hline
    J0459+0235A & 74.75059 & 2.59009 & 18.548 & \multirow{2}{*}{6.571} & \multirow{2}{*}{2025-09-26} & \multirow{2}{*}{NGPS} & \multirow{2}{*}{1800} & \multirow{2}{*}{136}\\
    J0459+0235B & 74.75187 & 2.58878 & 18.839 &  &  &  &  & \\
    \hline
    J0823+3503A & 125.97111 & 35.06128 & 19.646 & \multirow{2}{*}{3.234} & \multirow{2}{*}{2025-09-25} & \multirow{2}{*}{NGPS} & \multirow{2}{*}{1500} & \multirow{2}{*}{29}\\
    J0823+3503B & 125.97157 & 35.06210 & 18.709 &  &  &  &  & \\
    \hline
    J1601+3926A & 240.30114 & 39.44162 & 20.342 & \multirow{2}{*}{12.026} & - & - & - & \\
    J1601+3926B & 240.30446 & 39.44376 & 18.234 &  & 2021-04-17 & OMR & 3600 & 0\\
    \hline
    J2252+0847A & 343.15648 & 8.78938 & 22.801* & \multirow{2}{*}{8.650} & - & - & - & \\
    J2252+0847B & 343.15868 & 8.79041 & 17.269 &  & 2023-12-18 & BFOSC & 5400 & 0\\
    \hline
    J2312+0545A & 348.23114 & 5.76057 & 20.361 & \multirow{2}{*}{10.362} & \multirow{2}{*}{2024-01-10} & \multirow{2}{*}{YFOSC} & \multirow{2}{*}{4000} & \multirow{2}{*}{30}\\
    J2312+0545B & 348.23259 & 5.76306 & 19.131 &  &  &  &  & \\
    \hline
    J2314+2437A & 348.74984 & 24.62310 & 20.563 & \multirow{2}{*}{5.163} & \multirow{2}{*}{2022-09-26} & \multirow{2}{*}{BFOSC} & \multirow{2}{*}{7200} & \multirow{2}{*}{0}\\
    J2314+2437B & 348.74970 & 24.62453 & 19.137 &  &  &  &  & \\
    \hline
    \end{tabular}
    \tablefoot{* \textit{Gaia} DR3 had not public a reliable magnitude, so the DESI-LS DR10 $r$-band magnitude is provided. Columns (2) and (3) are \textit{Gaia} DR3 J2000 coordinates in degrees. Except for faint J2252+0847A, column (4) gives \textit{Gaia} DR3 $G$-band magnitude. Column (5) lists the angular separation between the two members in each pair calculated by their \textit{Gaia} DR3 J2000 coordinates. Column (8) lists the PA of the spectroscopic slit during the exposure, which is measured eastward from north.}
    \label{target_obs}
\end{table*}

\subsubsection{XLT216} \label{sec2.2.1}
The spectrograph made by Optomechanics Research Inc. (OMR) and the Beijing Faint Object Spectrograph and Camera (BFOSC) equipped on XLT216 \citep{FanZ2016XL216OMR-BFOSC} were used to take the spectra of two single targets (J1601+3926B and J2252+0847B) and a paired target (J2314+2437). They are both single-channel spectrographs. The OMR, with its 300 lines/mm grating blazed at 6\,000{\AA}, was chosen to take the spectrum of J1601+3926B on the night of 17 Apr. 2021. The seeing during the observation was approximately $2.0^{\prime\prime}$, and a 1.8{\arcsec}-wide slit was adopted, resulting in a dispersion of 4.0{\AA}/pixel. The actual wavelength coverage of this spectrum is from 3\,300{\AA} to 8\,900{\AA}. In addition, the BFOSC was chosen to take the spectra of a paired target (J2314+2437) and a single target (J2252+0847B) on the night of 26 Sep. 2022 and 18 Dec. 2023, respectively. The G4 grism and a slit with a width of 1.8{\arcsec} were chosen, yielding a dispersion of 4.45{\AA}/pixel. The actual wavelength coverage of these spectra is also from 3\,300{\AA} to 8\,900{\AA}.

\subsubsection{LJT240} \label{sec2.2.2}
The single-channel Yunnan Faint Object Spectrograph and Camera \citep[YFOSC,][]{WangCJ2019LJ240YFOSC} on LJT240 was used to take the spectra of 4 paired targets (J0229+0320, J0249+2606, J0428$-$0712, and J2312+0545) on the night of 15 Nov. 2022 and 10 Jan. 2024. The G3 grism (a dispersion of 2.9{\AA}/pixel) that covers the wavelength of 3\,400{\AA}--9\,100{\AA} and a slit with a width of 1.8{\arcsec} was chosen.

\subsubsection{P200} \label{sec2.2.3}
The dual-channel Double Spectrograph (DBSP) on P200 was used to take the spectra of 2 paired targets (J0130+0725 and J0242$-$1002) on the night of 16 Oct. 2023 (CTAP2023-B0014, P.I. Zizhao He). They are faint in \textit{Gaia} detection and have member separations around $\sim$ 2{\arcsec}. Benefiting from the large aperture of the P200 and the excellent observing conditions at Palomar Observatory, DBSP performs well when observing faint targets with small member separations. The dichroic D-55 was used to divide light into blue and red channels, and the exposure time for both channels was equal. The 300 lines/mm grating blazed at 3\,990{\AA} (a dispersion of 2.108{\AA}/pixel) was chosen for the blue channel, while the 316 lines/mm grating blazed at 7\,150{\AA} (a dispersion of 1.535{\AA}/pixel) was for the red channel \citep{Oke1982DBSP}. The seeing during the observation was approximately $1.5^{\prime\prime}$, so a $1.5^{\prime\prime}$-wide slit was chosen. Unfortunately, during the exposure of J0130+0725, the slit was rotated to an incorrect angle, leading to heavy blending of the spectra of the two members. In addition, there is a bad CCD region at the wavelength of approximately 5\,800{\AA}--6\,250{\AA} in the spectrum of the red channel. Thus, the spectra of the two paired targets are unavailable in this range.

Thanks to the newly operated Next Generation Palomar Spectrograph (NGPS) on P200, we obtained high-quality spectra of 2 paired targets (J0459+0235, J0823+3503) on the nights of 25 and 26 Sep. 2025 (CTAP2025-B0044, P.I. Zizhao He). The NGPS is designed with 4 spectral channels (U, G, R, and I), enabling simultaneous coverage of the 3\,100{\AA}--10\,400{\AA} wavelength range in a single exposure. During the 2025 NGPS observing run, only the R and I channels were available. The R channel covers 5\,610{\AA}--7\,940{\AA}, while the I channel covers 7\,560{\AA}--10\,400{\AA}. The grating of the NGPS is fixed, with R-channel dispersion of 0.57{\AA}/pixel and I-channel dispersion of 0.69{\AA}/pixel \citep{JiangHJ2018NGPS}. An on-chip binning setting\footnote{\url{https://www.astro.caltech.edu/documents/5328/NGPS-OBSERVE-USER-MANUAL.pdf}} allows observers to adjust spectral and spatial resolution, as well as the spectral SNR, to accommodate different science goals and observing conditions. This includes spectral binning, which adjusts the resolution along the dispersion direction, and spatial binning, which adjusts the resolution along the spatial direction. For observations of faint point sources, both binnings are recommended to be set to relatively large values to improve the spectral SNR, at the cost of spectral and spatial resolution. We adopted a spectral binning of 3 on both nights, yielding a dispersion of 1.71{\AA}/pixel in the R channel and 2.07{\AA}/pixel in the I channel. Spatial binning was set to 4 for the observations of J0459+0235, corresponding to a spatial scale of 1.0{\arcsec}/pixel in both channels. For J0823+3503, a binning of 2 was adopted, providing a spatial scale of 0.5{\arcsec}/pixel in both channels. The seeing was $\sim\ 1.5^{\prime\prime}$ on both nights, so a 1.5{\arcsec}-wide slit was chosen for both targets.

\subsection{Data reduction} \label{sec2.3}
The original spectra from the 5 spectrographs were reduced using a \textsc{Python}\footnote{\url{https://www.python.org/}} module \texttt{PySpecRedux} developed by us, which is dedicated to low-resolution long-slit spectra (Chen et al., in prep.). It was first used in \citet{HeZZ2025LeQDQPQ19}. The module employs several common modules such as \texttt{Astropy} \citep{AstropyCollaboration2022}, \texttt{NumPy} \citep{Harris2020NumPy}, \texttt{Pandas} \citep{McKinney2010Pandas}, and \texttt{SciPy} \citep{Virtanen2020SciPy} to implement standard spectral processing procedure. The procedure includes bias subtraction, flat-fielding, cosmic ray rejection \citep[\texttt{Astroscrappy}\footnote{\url{https://zenodo.org/record/1482019}}, ][]{McCully2018Astroscrappy}, background-subtraction, and aperture extraction, as well as wavelength and flux calibrations. All the reduced spectra were uploaded to an online interactive tool (Spectrum Viewer\footnote{\url{https://preview.lmytime.com/uspec/}}) to measure their redshifts. The redshift of each quasar was measured by averaging the redshifts of all major emission lines, with an accuracy of 3 decimal places. For DBSP and NGPS spectra, splicing was performed at $\sim$ 5\,500{\AA} and $\sim$ 7\,900{\AA}, respectively. The short- and long-wavelength ends of each spectrum were appropriately truncated where the spectra approach the CCD edges, exhibiting poor SNRs. Similar truncation was applied to the spectra from OMR, BFOSC, and YFOSC. All spectra displayed in this work were smoothed using the \textsc{Python} function \texttt{astropy.convolution.Gaussian1DKernel}\footnote{\url{https://docs.astropy.org/en/stable/api/astropy.convolution.Gaussian1DKernel.html}}, with both the standard deviation (\texttt{stddev}) and the kernel array size (\texttt{x\_size}) set to 5.
%--------------------------------------------------------------------

\section{Confirmation outcomes} \label{sec3}
It is insufficient to distinguish between QPs, LQs, and PQs by solely confirming both members in a pair as quasars based on their spectra. For QPs and PQs, it is generally necessary to estimate the radial velocity difference ($\lvert \Delta v_r \rvert$) between the two quasars from their spectroscopic redshifts. This allows one to determine whether they can be considered to lie at the same redshift and thus be in an interacting stage, or at least undergo a close encounter and produce mutual physical influence. A threshold of $\lvert \Delta v_r \rvert$ $\leqslant$ 2000 km/s is commonly adopted to distinguish QPs from PQs \citep[e.g.,][]{Hennawi2006QPdvr, Hennawi2010QPdvr}. According to \citet{Hogg1999QPdvr}, the $\lvert \Delta v_r \rvert$ between the two quasars was defined as:
\begin{equation}
    \lvert \Delta v_r \rvert = c\cdot\frac{\lvert z_{\rm A} - z_{\rm B} \rvert}{1 + (z_{\rm A} + z_{\rm B})/2},
\end{equation}
\label{Eq1}where $z_{\rm A}$ and $z_{\rm B}$ are redshifts of members A and B, respectively, and $c$ is the velocity of light. The threshold of $\lvert \Delta v_r \rvert \leqslant$ 2000 km/s confirmed 6 pairs with identical member redshifts and 3 PQs. By calculating the $r_p$ between the members of the 6 pairs, we found that none are below 17 kpc, indicating that the 6 pairs are either DQs or potential LQs, rather than BQs that are separated only at the pc-scale. For DQs and LQs, both spectroscopic and imaging data are generally required to distinguish between them. In the vast majority of DQs, the spectral features of the member quasars are distinctly different, particularly in terms of emission-line profiles, continuum shapes, and the presence or absence of intrinsic broad absorption lines \citep[e.g.,][]{Farina2011lowZ6QP, Shields2012BQLBQS0103-2753, Altamura2020SerendipitousDQ, ChenYC2021DECamLSDQLQ, YueMH2021kpcQPJ2037, Koss2023DAGN230pcUGC4211}. In some cases, they may also exhibit signs of interaction between their host galaxies \citep[e.g.,][]{LiuX2011LowZAGNPair, ZhuYD2024QGpairALMAimaging, DengZJ2026LowZAGNPair, YueMH2026CloseQPJ2037}. In contrast, LQs often have highly similar spectra and show a detectable foreground massive deflector projected between them in imaging data \citep[e.g.,][]{Lemon2024GLreview, Ducourant2026GaiaGraL-X}.

\begin{figure*}
    \centering
    \includegraphics[width=\textwidth]{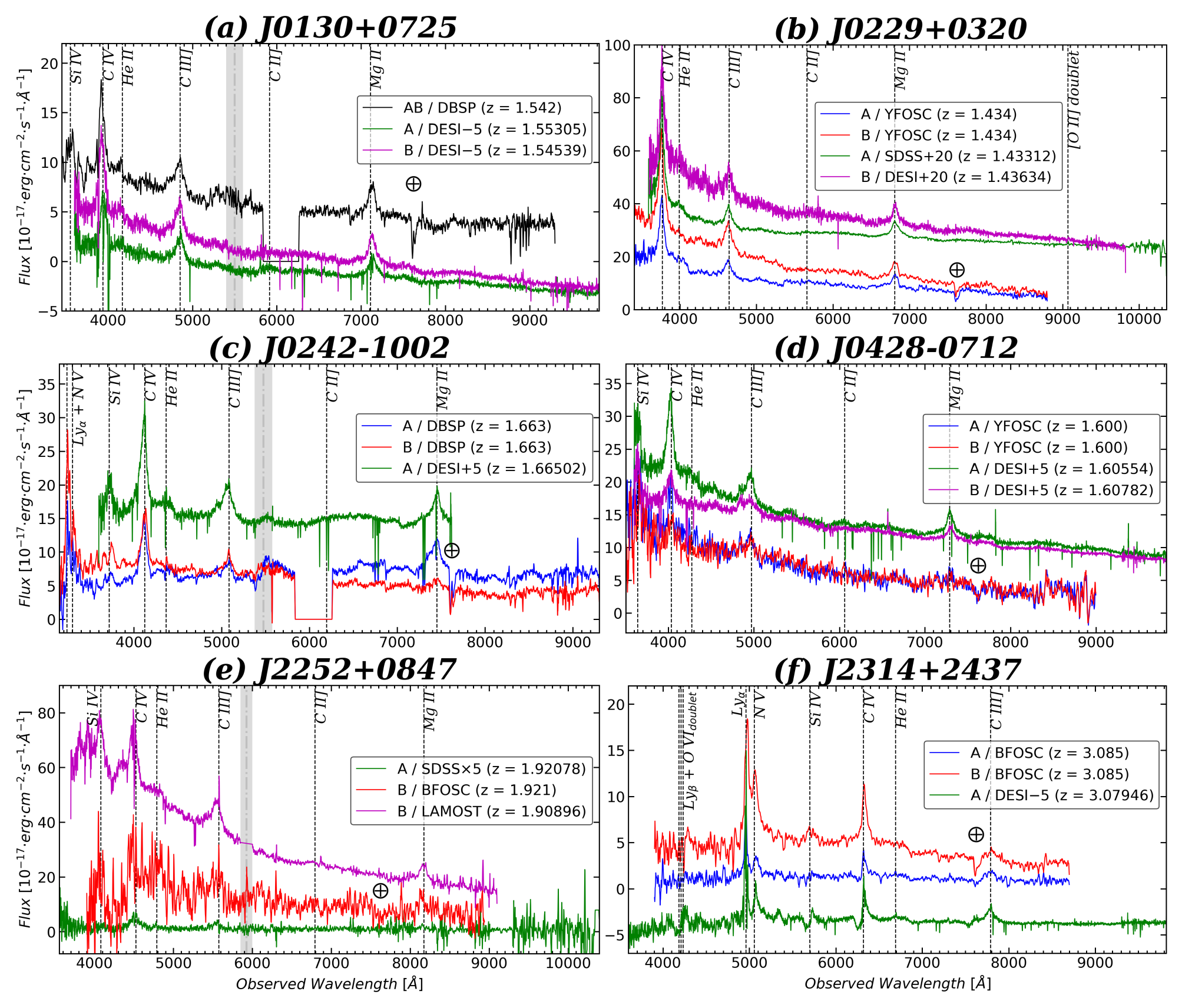}
    \caption{Spectra of the 6 confirmed DQs. Unfortunately, the slit of J0130+0725 was rotated to an incorrect angle, and the spectra of members A and B were heavily blended. The grey regions in the spectra of J0130+0725, J0242-1002, and J2252+0847 denote the splicing of spectra from two channels of DBSP and LAMOST. Each black open circle with a cross denotes telluric absorption at $\sim$ 7\,600{\AA}. All marked emission lines are attributed to quasar emission. Public spectra plotted in each panel are detailed as follows. Panel (a): The spectra of members A and B are from DESI DR1 observations on MJD 59524 and MJD 59501, respectively; panel (b): SDSS DR17 observation on MJD 58097 for A, while DESI DR1 observation on MJD 59569 for B; panel (c): DESI DR1 observation on MJD 59552 for A; panel (d): DESI DR1 observations on MJD 59527 and MJD 59504 for A and B, respectively; panel (e): SDSS DR17 observation on MJD 55889 for A, while LAMOST DR13 observation on MJD 57327 for B; panel (f): DESI DR1 observation on MJD 58097 for A. The SDSS DR17 spectral flux of J2252+0847A is magnified by a factor of 5 to exhibit its emission features clearly. The other public spectra  (except for the LAMOST DR13 spectrum of J2252+0847B) are plotted with their fluxes shifted vertically by $-$5--20 units to avoid overlap. The observed flux of J2252+0847B with BFOSC is underestimated due to the poor seeing of $\sim$ 3{\arcsec} and the adopted 1.8{\arcsec}-wide slit.}
    \label{DQ_Spectra}
\end{figure*}

\begin{figure}
    \centering
    \includegraphics[width=0.5\textwidth]{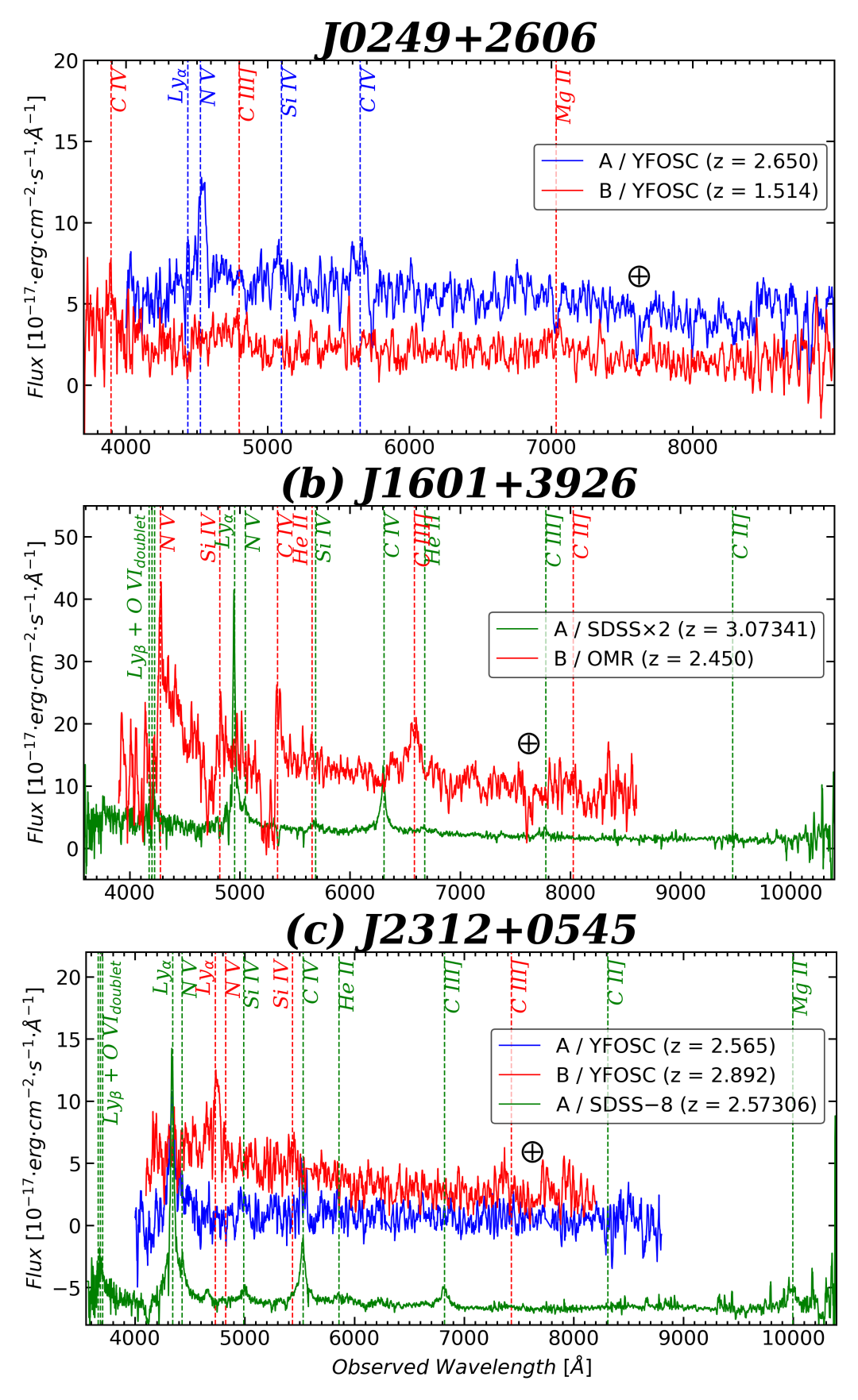}
    \caption{Similar to Figure \ref{DQ_Spectra} but for the 3 confirmed PQs. Public spectra plotted in each panel are detailed as follows. J1601+3926 panel: SDSS DR17 observation on MJD 56066 for A; J2312+0545 panel: SDSS DR17 observation on MJD 55883 for A. The SDSS DR17 spectral flux of J1601+3926A is magnified by a factor of 2 to exhibit its emission features clearly. The SDSS DR17 spectral flux of J2312+0545A is shifted by $-$8 units to avoid overlap.}
    \label{PQ_Spectra}
\end{figure}

\begin{figure*}
    \centering
    \includegraphics[width=\textwidth]{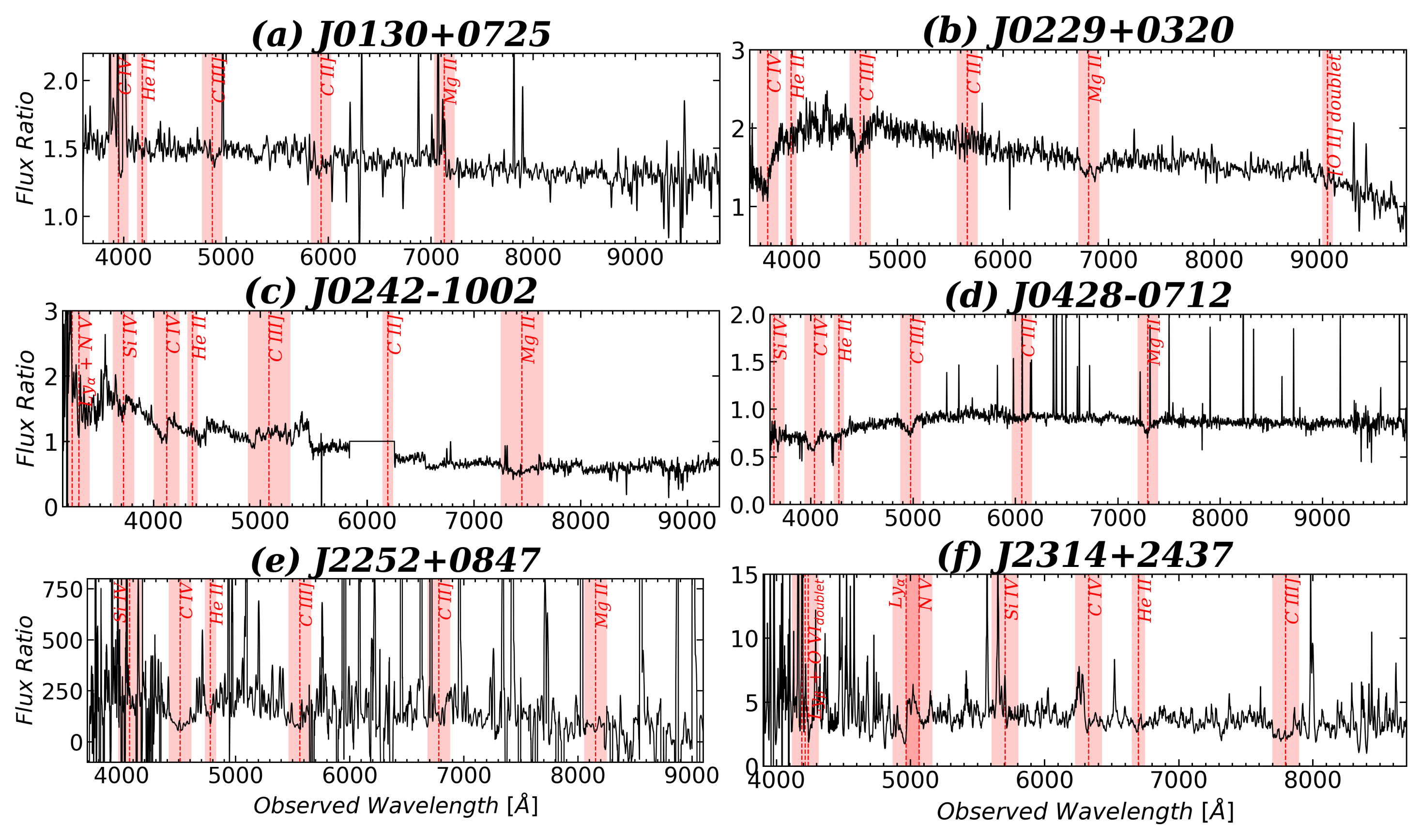}
    \caption{Spectral flux ratio curves of 6 DQs. All flux ratios were obtained by dividing the spectrum of member B by that of member A, and were smoothed with appropriate Gaussian kernel parameters for visual clarity. The shaded red regions in each panel highlight the differences in the flux ratio around the emission lines. The flux ratios for J0130+0725, J0229+0320, J0428$-$0712, and J2252+0847 were obtained from the public spectra of the two members, selecting those with the closest observation dates to maximize quasi-simultaneity and minimize spectral differences introduced by quasar variability. The flux ratio jump of J0242$-$1002 at $\sim$ 4\,500{\AA} is due to the bump edge of the Fe II pseudo-continuum, while that at $\sim$ 5\,450{\AA} is due to the spectral splicing.}
    \label{SpecFR}
\end{figure*}

During our 5-year spectroscopic follow-up campaign, SDSS and DESI released new spectroscopic data \citep[e.g., SDSS DR17 and DESI DR1,][]{Abdurrouf2022SDSSDR17, DESICollaboration2025DESIDR1}, such that nearly all targets reported in this work obtained SDSS or DESI spectra. This presented both challenges and opportunities for confirming DQs in the MGQPC. We adopted these public spectra as auxiliary data to improve our DQ confirmation. Spectra from SDSS and DESI were retrieved via SPARCL \citep{Juneau2025SPARCL}. The public spectrum of J2252+0847B was retrieved from the newest LAMOST DR13\footnote{\url{https://www.lamost.org/dr13/v1.0/search}} \citep{Cui2012LAMOST}. After comparing the member spectra in each pair and examining the features in the DESI-LS DR10 images, we conclude that all the 6 pairs with identical member redshifts are DQs. The remaining two, J0459+0235 and J0823+3503, are both quasar-star projections (QSPs). The confirmation outcomes of all the 11 candidate pairs are summarized in Table \ref{confirm_outcomes}, where the redshifts (rounded to 5 decimal places) are updated with the help of public spectra. The spectra of the 6 DQs and 3 PQs are shown in Figure \ref{DQ_Spectra} and \ref{PQ_Spectra}, respectively. In addition, the spectral flux ratios of the 6 DQs were plotted in Figure \ref{SpecFR} to further examine the differences in their spectral features. The following subsections describe the 6 DQs and 3 PQs, respectively. The spectra of two QSPs are shown in Figure \ref{QS_Spectra}. The relevant brief description and discussion are presented in Appendix \ref{appendixA} and Section \ref{sec4}, respectively.

\begin{table*}
    \caption{The spectroscopic confirmation outcomes of the 11 observed candidates.}
    \centering
    \begin{tabular}{c|c|c|c|c|c|c}
    \hline \hline
    Name$^{\,(1)}$ & Redshift$^{\,(2)}$ & SepAB$^{\,(3)}$ & $\overline{r_p}^{\,(4)}$ & $\lvert \Delta v_r \rvert^{\,(5)}$ & Source Type$^{\,(6)}$ & Outcome$^{\,(7)}$\\ &  & ({\arcsec}) & ($kpc$) & ($km\ \cdot\ s^{-1}$) &  & \\
    \hline
    J0130+0725A & 1.55305 & \multirow{2}{*}{2.051} & \multirow{2}{*}{17.372} & \multirow{2}{*}{900.829} & quasar & \multirow{2}{*}{DQ} \\
    J0130+0725B & 1.54539 &  &  &  & quasar &  \\
    \hline
    J0229+0320A & 1.43312 & \multirow{2}{*}{2.086} & \multirow{2}{*}{17.614} & \multirow{2}{*}{396.484} & quasar & \multirow{2}{*}{DQ} \\
    J0229+0320B & 1.43634 &  &  &  & quasar &  \\
    \hline
    J0242$-$1002A & 1.66502 & \multirow{2}{*}{2.384} & \multirow{2}{*}{20.194} & \multirow{2}{*}{227.319} & quasar & \multirow{2}{*}{DQ} \\
    J0242$-$1002B & 1.663 &  &  &  & quasar &  \\
    \hline
    J0249+2606A & 1.514 & \multirow{2}{*}{3.210} & \multirow{2}{*}{27.175} & \multirow{2}{*}{110\,501.049} & quasar & \multirow{2}{*}{PQ} \\
    J0249+2606B & 2.650 &  &  &  & quasar &  \\
    \hline
    J0428$-$0712A & 1.60554 & \multirow{2}{*}{4.042} & \multirow{2}{*}{34.247} & \multirow{2}{*}{262.221} & quasar & \multirow{2}{*}{DQ} \\
    J0428$-$0712B & 1.60782 &  &  &  & quasar &  \\
    \hline
    J0459+0235A & 1.420 & \multirow{2}{*}{6.571} & \multirow{2}{*}{-} & \multirow{2}{*}{-} & quasar & \multirow{2}{*}{Q+S} \\
    J0459+0235B & 0.000 &  &  &  & star &  \\
    \hline
    J0823+3503A & 0.79764 & \multirow{2}{*}{3.234} & \multirow{2}{*}{-} & \multirow{2}{*}{-} & quasar & \multirow{2}{*}{Q+S} \\
    J0823+3503B & 0.000 &  &  &  & star &  \\
    \hline
    J1601+3926A & 3.07341 & \multirow{2}{*}{12.026} & \multirow{2}{*}{97.489} & \multirow{2}{*}{49\,683.220} & quasar & \multirow{2}{*}{PQ} \\
    J1601+3926B & 2.450 &  &  &  & BAL quasar &  \\
    \hline
    J2252+0847A & 1.92078 & \multirow{2}{*}{8.650} & \multirow{2}{*}{72.722} & \multirow{2}{*}{1\,215.679} & quasar & \multirow{2}{*}{DQ} \\
    J2252+0847B & 1.90896 &  &  &  & quasar &  \\
    \hline
    J2312+0545A & 2.57306 & \multirow{2}{*}{10.362} & \multirow{2}{*}{83.113} & \multirow{2}{*}{25\,616.889} & quasar & \multirow{2}{*}{PQ} \\
    J2312+0545B & 2.892 & &  &  & quasar &  \\
    \hline
    J2314+2437A & 3.07946 & \multirow{2}{*}{5.163} & \multirow{2}{*}{39.444} & \multirow{2}{*}{406.849} & BAL quasar & \multirow{2}{*}{DQ} \\
    J2314+2437B & 3.085 &  &  &  & quasar &  \\
    \hline
    \end{tabular}
    \tablefoot{Column (2) lists the redshifts of the members in each pair. Redshift values that are rounded to 5 decimal places are updated by public spectra. Column (3) is the angular separation between the two members. For the DQ, column (4) is the average $r_p$ calculated with the member separation and average redshift of the two member quasars. For the PQ, it represents the projected distance calculated with the member separation and the redshift of the foreground quasar. Column (5) is the radial velocity difference between the two quasars, which is an indicator to distinguish DQ or LQ from PQ.}
    \label{confirm_outcomes}
\end{table*}

\subsection{Dual quasars} \label{sec3.1}
The combination of imaging features and spectral differences can, to a certain extent, effectively confirm DQs. In terms of imaging features, the clear color difference between the two quasars and the absence of a foreground deflector are key indicators. Together with spectral differences (e.g., emission-line profiles, continuum shapes, and the presence or absence of intrinsic broad absorption lines), one can determine whether a system is a DQ rather than an LQ. The spectral flux ratio provides a relatively rapid way to distinguish spectral differences compared to spectral fitting, but it requires that the spectra of the two quasars be taken simultaneously or quasi-simultaneously. The slit rotation in our follow-up observations was adopted for this purpose. However, for 4 of 6 DQs, either the spectral quality and resolution were insufficient in our observation, or the slit was not rotated to the required angle, leaving us no choice but to obtain their flux ratios from public spectra with the closest observation dates.

\subsubsection{J0130+0725} \label{sec3.1.1}
This pair was firstly selected as an LQ candidate in both \citet[][hereafter, Rusu19]{Rusu2019LQCQPCinPS1J0130J0229J0242J0249} and \citet{HeZZ2023LScolorLQC}, but was observed and reported as an ``unclassified quasar pair'' in \citet{Lemon2023GLQG150LeQ+QP+PQPJ0130J0249}. The incorrect slit rotation caused the DBSP spectra of the two quasars to overlap, making them spatially unresolved and hindering a robust identification of this pair (see panel (a) of Figure \ref{DQ_Spectra}). DESI DR1 spectra indicate that both members are quasars with the same redshifts ($\lvert \Delta v_r \rvert \approx$ 900 km/s) and have nearly identical spectra. To further confirm whether this pair is a DQ, the two public spectra with the closest observation dates -- namely, the DESI DR1 spectra of J0130+0725A (MJD 59524) and J0130+0725B (MJD 59501) -- were used to obtain the spectral flux ratio. The flux ratio curves around the emission lines C IV, C III], and Mg II show mild fluctuations, particularly around C III] and Mg II (see panel (a) of Figure \ref{SpecFR}). This suggests that the two members might be intrinsically different quasars rather than two lensed images of the same background quasar. Furthermore, no detectable foreground deflector is found in the imaging data of this pair. We therefore consider J0130+0725 a DQ at $z\ \sim$ 1.55 and separated by an average transverse distance $\overline{r_p}$ of 17.372 kpc, which is the smallest among the DQs in this work.

Although the observational features support the DQ interpretation of J0130+0725, the LQ hypothesis cannot be completely ruled out. The slight emission-line differences between the two quasars could be due to the gravitational microlensing effect caused by the outskirts of the foreground deflector \citep[e.g.,][]{Sluse2012MicroLQBLR, Fian2021MicroLQBLR, Hutsemekers2023MicroLQJ1004+4112BLR}, or due to the time-delayed variability of the background quasar caused by the strong lensing effect. In addition, the imaging data show that the two quasars have very similar colors, and an undetected foreground deflector might be too faint and outshone by the light of the quasar images. In future work of this pair, lens-based modeling combined with high-resolution deep imaging, or light-curve analysis of the two quasars, will be needed to assess the LQ hypothesis definitively. Alternatively, multi-band observations from $\gamma$-ray to radio are required to confirm that the two quasars are physically distinct.

\subsubsection{J0229+0320} \label{sec3.1.2}
This pair was also selected as an LQ candidate in Rusu19, and was first reported and briefly described in \citet[][hereafter, Lemon19]{Lemon2019PhDJ0229J0249} and \citet[][hereafter, Lemon20]{Lemon2020STRIDES10LQ+10QPJ0229} as a DQ or nearly identical quasar \citep[NIQ, e.g.,][]{Anguita2018STRIDES-II-2LQ4PQ7NIQ, Lemon2018GLQG-II24LQ, Lemon2020STRIDES10LQ+10QPJ0229, Lemon2023GLQG150LeQ+QP+PQPJ0130J0249, Hawkins2021LQ2138-431Galaxy, ChenYH2023DMLQC}. The spectra of J0229+0320A and B are highly similar (see panel (b) of Figure \ref{DQ_Spectra}), and their colors in the imaging data are also nearly identical. The spectral flux ratio was obtained from the SDSS DR17 spectrum of J0229+0320A (MJD 58097) and the DESI DR1 spectrum of J0229+0320B (MJD 59569). It shows apparent fluctuations around the broad emission lines C IV, C III], and Mg II, indicating that the two quasars would be intrinsically different (see panel (b) of Figure \ref{SpecFR}). In Lemon20, the spectral flux ratio of J0229+0320 obtained by simultaneous observation of member A and B also reveals an apparent fluctuation around the Mg II emission line (double-peak structure; see Figure 4 in Lemon20).

Despite the apparent fluctuations in flux ratio around these emission lines in Lemon20 and this work, J0229+0320 could also be an LQ that appears as an NIQ. Time-delayed variability of the background quasar and microlensing effects can also distort the emission-line profiles, thereby leading to fluctuations in the flux ratio. Lemon19 also provided the DES $i$-band light curves of the two quasars, and regarded the variability of them to be similar, thus keeping the possibility of J0229+0320 to be an LQ. However, neither Lemon20 nor this work detected a reliable foreground deflector between them in the imaging data. The DES $i$-band light curves in Lemon19 contain only 4 observations for each member quasar, making it impossible to determine whether the variability between the two quasars exhibits a reliable time delay induced by gravitational lensing. This time delay could be a powerful observational piece of evidence supporting an LQ. Unless the host galaxy of the background quasar distorted by strong lensing is detected in high-resolution imaging data (appears as arcs across the quasar images), or a reliable time delay is derived from high-quality light curves, we have little choice but to consider J0229+0320 a DQ at $z\ \sim$ 1.55 with a $\overline{r_p}$ of 17.614 kpc.

\subsubsection{J0242$-$1002} \label{sec3.1.3}
Rusu19 selected this system as another LQ candidate. Both the DBSP and DESI DR1 spectra of J0242$-$1002A reveal a bump of Fe II pseudo-continuum around the He II emission line, which is absent in the DBSP spectrum of J0242$-$1002B. In addition, a sub-peak is present in the blue wing of C III] in the two spectra of J0242$-$1002A, but is not seen in that of J0242$-$1002B (see panel (c) of Figure \ref{DQ_Spectra}). The Mg II line intensity ratio is notably different from those of C IV and C III] emission lines, which is unlikely to be caused by gravitational lensing. These spectral differences indicate that the two quasars are intrinsically distinct. The spectral flux ratio of the two quasars was obtained from our DBSP spectra. It shows apparent fluctuations around Si IV, C IV, C III], and Mg II (see panel (c) of Figure \ref{SpecFR}). Moreover, no detectable foreground deflector is found in the imaging data. Therefore, we conclude that J0242$-$1002 is a DQ at $z\ \sim$ 1.66 with a $\overline{r_p}$ of 20.194 kpc.

\subsubsection{J0428$-$0712} \label{sec3.1.4}
The YFOSC spectra of the two quasars in this pair exhibit nearly identical continuum shapes and emission-line profiles. Due to the low spectral SNR and poor data quality caused by the poor seeing during the exposure, reliable spectral differences are difficult to discern. DESI DR1 spectra provide clear spectral features for J0428$-$0712 (see panel (d) of Figure \ref{DQ_Spectra}). The spectral flux ratio was obtained from the DESI DR1 spectra of J0428$-$0712A (MJD 59527) and J0428$-$0712B (MJD 59504). It reveals that the C IV, C III], and Mg II lines of J0428$-$0712A are significantly different from those of J0428$-$0712B, which suggests that the two members are intrinsically different quasars rather than a doubly imaged LQ (see panel (d) of Figure \ref{SpecFR}). Along with the non-detection of a foreground deflector in the imaging data, we consider J0428$-$0712 a DQ at $z\ \sim$ 1.61 with a $\overline{r_p}$ of 34.247 kpc.

Similar to J0130+0725 and J0229+0320, the possibility of J0428$-$0712 being an LQ cannot be completely ruled out. Although no reliable foreground deflector is detected between them, the spectroscopic and imaging data show that their colors are nearly identical, and the flux ratio fluctuations around the emission lines could also be caused by time-delayed quasar variability and microlensing effects. We believe this pair is a DQ, but high-resolution imaging or high-quality light curves are required to fully rule out the LQ hypothesis.

\subsubsection{J2252+0847} \label{sec3.1.5}
This pair exhibits a large difference in brightness between its two members. Because J2252+0847A is too faint (with a DESI-LS DR10 $r$-band magnitude of 22.8) and SDSS DR17 has already published its scientific spectrum, we observed only the spectrum of J2252+0847B (see panel (e) of Figure \ref{DQ_Spectra}). However, the seeing was as poor as $\sim$ 3{\arcsec} during the exposure, and the adopted 1.8{\arcsec}-wide slit was too narrow, resulting in an underestimation of the observed flux of J2252+0847B and an unreliable redshift measurement. The spectral flux ratio was obtained from the SDSS DR17 spectrum of J2252+0847A (MJD 55889) and the LAMOST DR13 spectrum of J2252+0847B (MJD 57327). Although the flux ratio has a poor signal-to-noise ratio, the fluctuations around C IV and C III] are apparent (see panel (e) of Figure \ref{SpecFR}), suggesting the intrinsic distinction of the two quasars. In the imaging data, no reliable foreground deflector is detected between the two quasars, and their large brightness difference and angular separation do not seem to support the LQ hypothesis. Although the redshift difference between the two quasars is $\sim$ 0.012, which is the largest among the 6 DQs, their $\lvert \Delta v_r \rvert$ still meets our criterion. We therefore conclude that J2252+0847 is a DQ at $z\ \sim$ 1.91 with a $\overline{r_p}$ of 72.722 kpc.

\subsubsection{J2314+2437} \label{sec3.1.6}
This pair is an evident DQ and has the highest redshift among the observed pairs in this work. Both the Si IV and C IV emissions of J2314+2437A exhibit broad absorption line features \citep[BALs; e.g.,][]{WangFG2018z7BALQSO, YiWM2021MgIIBAL} in their blue wings, which are attributed to intrinsic broad absorptions, i.e., fast outflows \citep[e.g.,][]{Elvis2000BALQSOmodel, FuXD2023PG1001+054FastOutflow, Naddaf2023BALQSOmodel}. These features are clearer in its DESI DR1 spectrum (MJD 58097), whereas J2314+2437B shows no such features (see panel (f) of Figure \ref{DQ_Spectra}). The spectral flux ratio of the two quasars, obtained from our BFOSC spectra, indicates that Ly$\alpha$, N V, Si IV, C IV, and C III] of both quasars all exhibit varying degrees of difference (see panel (f) of Figure \ref{SpecFR}). The presence of BAL features in Si IV and C IV in J2314+2437A and their absence in J2314+2437B robustly rule out the possibility that J2314+2437 is an LQ. In the imaging data, the undetected foreground deflector between them further confirms that J2314+2437 is a DQ at $z\ \sim$ 3.08 with a $\overline{r_p}$ of 39.444 kpc.

\subsection{Projected quasars} \label{sec3.2}
PQs refer to systems in which two quasars appear close in projection but are physically unassociated and located at different cosmological distances \citep[e.g.,][]{Heintz2016PQHAQ2358+1030, Lau2017QPQ-PhD, Lau2018QPQ-IX}. Despite the lack of dynamical association, PQs hold unique value in the study of the intergalactic and circumgalactic medium, providing critical observational constraints on the cycling and evolution of baryonic matter in the universe \citep[e.g.,][]{Hennawi2006QPQ-I, Prochaska2009QPQ-III, Hennawi2013QPQ-IV, Farina2014PQCGM, ChenZF2023PQCGM, FanXH2023QuasarIGM}. PQs are much easier to identify compared to DQs and LQs. A large redshift difference and a $\lvert \Delta v_r \rvert$ far exceeding 2000 km/s can quickly filter them out.

In Rusu19, J0249+2606 was selected as an LQ candidate based on color similarity and visual inspection, but was observed and confirmed as a PQ in \citet{Lemon2023GLQG150LeQ+QP+PQPJ0130J0249}. It is a system consisting of a foreground quasar at $z$ = 1.514 and a background quasar at $z$ = 2.650. Both spectra are displayed in the top panel of Figure \ref{PQ_Spectra}. For J1601+3926, since SDSS DR17 had already published the spectrum of J1601+3926A (MJD 56066), we observed only the spectrum of J1601+3926B. A foreground quasar at $z$ = 3.073 and a background quasar at $z$ = 2.450 constitute this PQ. In the OMR spectrum of J1601+3926B, Ly$\alpha$ is heavily absorbed, and both Si IV and C IV exhibit prominent BAL features in their blue wings (see middle panel of Figure \ref{PQ_Spectra}), indicating that it is a BAL quasar. J2312+0545 was selected as a wide-separation lensed quasar \citep[WSLQ,][]{Inada2003WSLQJ1004+4112, Inada2006WSLQJ1029+2623, Dahle2013WSLQJ2222+2745, ShuYP2018WSLQJ0909+4449, ShuYP2019WSLQJ1326+4806, Stern2021GraL-VIWSLQJ1651-0417DragonKite, Martinez2023COOLLAMPS-WSLQJ0542-2125, Napier2023COOLLAMPS-WSLQJ0335-1927} candidate in \textsc{Paper-I} due to its large separation ($\sim$ 10.4{\arcsec}), LQ-like configuration in imaging data, and similar spectral energy distributions (SEDs) between the two members. Our YFOSC spectra reveal that the redshift difference between the two quasars is only $\sim$ 0.32, the smallest among the PQs in this work (see bottom panel of Figure \ref{PQ_Spectra}). The red galaxy located between the two quasars is faint, and its YFOSC spectrum could hardly be extracted, making it impossible to measure its redshift. Such a wide-separation projection system that imitates an LQ-like configuration may hinder the discovery of DQs and WSLQs. We discuss this issue in Section \ref{sec4.2}.
%--------------------------------------------------------------------

\section{Discussion} \label{sec4}
We discuss 3 main topics aimed at optimizing the QP confirmation strategy and guiding future spectroscopic follow-up and identification efforts. The 3 topics include analyzing the color features of the observed targets, using photometric redshift estimation to improve the selection of QP candidates, and future confirmation strategies that could provide decisive observational evidence.

\subsection{Color feature} \label{sec4.1}
\begin{figure*}
    \centering
    \includegraphics[width=0.8\textwidth]{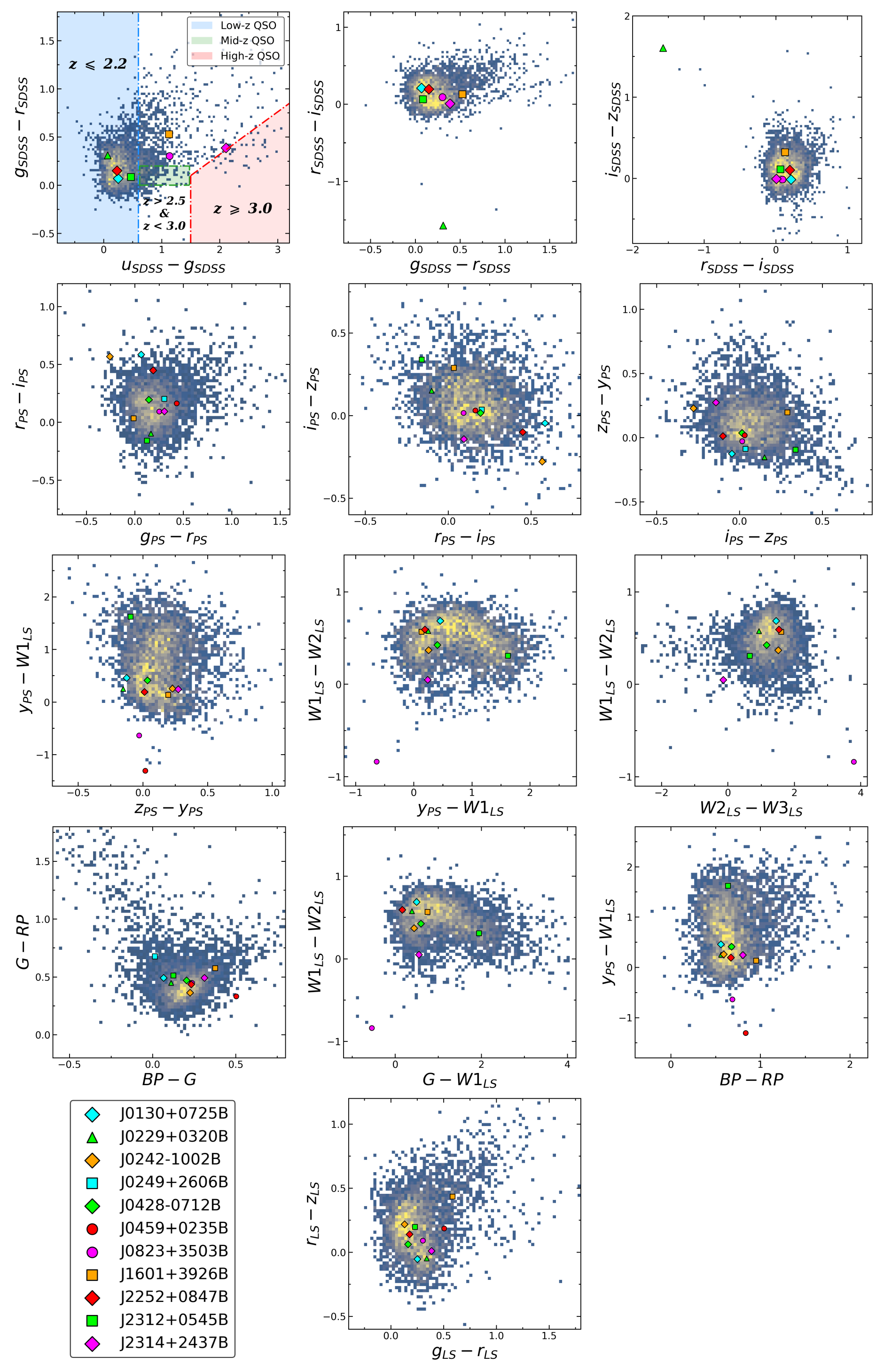}
    \caption{Color features of member B of the 11 observed targets. A subset of quasars was overlaid on each diagram as a 2D histogram with a yellow-to-blue gradient (high to low density), and the number of quasars in each diagram is equal to 4\,000. The legends for targets are in the lower-left corner. All magnitudes are given in the AB system. The blue, green, and red areas in the top left panel displayed the SDSS quasar selection for low-redshift ($z \leqslant 2.2$), mid-redshift ($2.5 < z < 3.0$), and high-redshift ($z \geqslant 3$) quasars, respectively. Note that the color of J0229+0320B may be problematic in the SDSS $r$-, $i$-, and $z$-band photometry (color abnormality in the last two panels of the first row).}
    \label{TargColors}
\end{figure*}

To evaluate the color features of member B of the 11 observed candidate pairs, 13 types of color-color diagrams were plotted in Figure \ref{TargColors}, 9 of which are the same as those in Figure 11 of \textsc{Paper-I}. These color diagrams incorporate multi-band photometric data from SDSS DR16 ($u_{\rm SDSS}$, $g_{\rm SDSS}$, $r_{\rm SDSS}$, $i_{\rm SDSS}$, $z_{\rm SDSS}$), PanSTARRS DR1 ($g_{\rm PS}$, $r_{\rm PS}$, $i_{\rm PS}$, $z_{\rm PS}$, $y_{\rm PS}$), DESI-LS DR10 ($g_{\rm LS}$, $r_{\rm LS}$, $z_{\rm LS}$, $W1_{\rm LS}$, $W2_{\rm LS}$, and $W3_{\rm LS}$), and \textit{Gaia} ($G$, $BP$, and $RP$). In addition, a subset of 4\,000 quasars was overlaid on each diagram (2D histogram with a yellow–blue color scale). These quasars were randomly selected from \texttt{GoodQSO\_Mag} in \textsc{Paper-I} and have all the aforementioned multi-band photometric data. Because J0242$-$1002, J0249+2606, J0428$-$0712, and J0459+0235 are not within the SDSS DR16 sky coverage, they are absent from the SDSS color diagrams (first row in Figure \ref{TargColors}). J0249+2606 is not within the DESI-LS DR10 sky coverage, and is therefore absent from the color diagrams that include DESI-LS photometric bands. In addition, DESI-LS DR10 does not provide a $W2$ magnitude for J0459+0235B, so it is absent from the color diagrams that include $W2_{\rm LS}$.

In these color diagrams, most of the confirmed quasars in this work are consistent with the quasar distribution, but two stars (J0823+3503B and J0459+0235B) are also within the distribution. This indicates that although color-based quasar selection works well, a small level of stellar contamination remains. On the other hand, a few genuine quasars that imitate stellar colors are excluded by the color method, such as J2314+2437B. As an example, the SDSS $ugr$ color diagram (top left panel in Figure \ref{TargColors}) overlaid with the criteria region of SDSS quasar selection \citep[blue area for $z \leqslant 2.2$, green for $2.5 < z < 3.0$, and red for $z \geqslant 3$; see Figure 7 in][]{Richards2002SDSSquasarColorSelection} was used to explain why the SDSS spectroscopic survey missed J1601+3926B and J2314+2437B. In this color diagram, J2314+2437B has $u-g = 2.107$ and $g-r = 0.388$. The quasar selection boundary for $z \geqslant 3$ at $u-g = 2.107$ gives $g-r \sim 0.367$. This indicates that J2314+2437B lies just outside the high-$z$ quasar selection boundaries and was therefore omitted by the SDSS spectroscopic surveys. The redshift of J1601+3926B indicates that it resides in the so-called ``redshift desert'' of SDSS quasars \citep[z $\sim$ 2.2–3; e.g.,][]{WuXB2010RedshiftDesert1, WuXB2010RedshiftDesert2} and was previously not selected as a quasar candidate by the SDSS spectroscopic survey. Quasars in the redshift desert are difficult to isolate from the crowded stars in color diagrams. A large amount of stellar contamination, such as J0823+3503B, may seriously drop the purity of quasar selection. The infrared color method may be a relatively effective way to remove such stellar contamination. In the color diagrams that include $W1_{\rm LS}$, the two stars J0823+3503B and J0459+0235B appear to be easily removable by a simple infrared color cut \citep[e.g.,][]{Stern2012AGNselectionMIR, Mateos2012AGNselectionWISE-IR, Assef2018WISEAGNcatalog}. However, even a combination of optical and infrared color methods may still not be able to effectively remove stars that closely imitate the colors of quasars \citep[e.g.,][]{Chaussidon2023DESIQSOselection}. Therefore, the astrometric method of isolating quasar candidates and its advantage over the color method presented in \textsc{Paper-I} can effectively recover quasars missed by the color method.

Although J0459+0235 is within the sky coverage of DESI-LS DR10, it is not within that of DR9, which defines the footprint of the DESI spectroscopic survey. Therefore, it has no DESI DR1 spectrum and remains unidentified previously. For candidates such as J0249+2606 and J0459+0235 that lie outside the footprints of current large-scale optical spectroscopic surveys (e.g., SDSS, DESI, and LAMOST), high-purity quasar selection methods and future spectroscopic follow-up remain necessary.

\subsection{Photometric redshift} \label{sec4.2}
This subsection discusses the importance of photometric redshift (photo-$z$) estimation in mitigating contamination from PQs and QSPs. In \textsc{Paper II}, the photo-$z$ values of member B (photo-$z_{\rm B}$) of the MGQPCs were meticulously estimated. Due to a lack of necessary photometric data for some member B sources, their photo-$z_{\rm B}$ values are absent. Among the 11 observed targets in this work, only 5 pairs have photo-$z_{\rm B}$ estimation. They are J0130+0725B with a photo-$z_{\rm B}$/spectroscopic redshift (spec-$z_{\rm B}$) of 1.583/ 1.545, J0242$-$1002B with 1.744/1.663, J1601+3926B with 2.492/2.450, J2312+0545B with 0.708/2.892, and J2314+2437B with 3.203/3.085. Except for J2312+0545B, the deviation between photo-$z_{\rm B}$ and spec-$z_{\rm B}$ ($\Delta z = \lvert$spec-$z_{\rm B}$ $-$ photo-$z_{\rm B}$$\rvert$) is within 0.04--0.12, yeilding the ratio $\Delta z$/spec-$z_{\rm B}$ ranges in $\sim$ 1.7\,\%--4.9\,\%. Although the values of $\lvert \Delta v_r \rvert$ derived from their photo-$z_{\rm B}$ and spec-$z_{\rm A}$ are all greater than 2000 km/s, indicating that they are PQ candidates, the small $\Delta z$ and $\Delta z$/spec-$z_{\rm B}$ are sufficient to place them at high priority for spectroscopic follow-up observations. This could improve the success rate of spectroscopic confirmation of DQs to some extent.

Removing PQ and QSP contamination using reliable photo-$z$ estimation is effective, even though these systems imitate LQ-like or WSLQ-like configurations (e.g., J0459+0235, J0823+3503, and J2312+0545). Currently, there are more than one hundred such potential contaminants still in the MGQPC catalog. Without further filtering, this could significantly drop the success rate of future spectroscopic follow-up and thus hinder the discovery of DQs and WSLQs. It is necessary to apply auxiliary quasar selection methods (e.g., SED similarity, precise photo-$z$ estimation) to the MGQPC catalog to reduce PQ and QSP contamination and thereby enable efficient spectroscopic follow-up for QP candidates with LQ-like configurations. For pair systems like J2312+0545, although they ultimately turn out not to be DQs or LQs, they still play an important role as probes of the CGM at different locations of the foreground galaxy \citep[e.g.,][]{Krishnarao2022MagellanicCGM}. Therefore, their discovery and confirmation may still prove beneficial.

\subsection{Future confirmation strategy}
The most fundamental aspect of selecting DQ candidates is to minimize stellar contamination. When neither color-based nor astrometric methods are sufficient to reduce stellar contamination, non-optical or other indirect methods become necessary. Examples include searching for $\gamma$-ray, X-ray, submillimeter, and radio imaging data of the candidates. In these electromagnetic bands, quasars still exhibit strong emission, whereas stars produce little or only very short-lived emission. In addition, quasars and stars can also be distinguished through variability \citep[e.g.,][]{Butler2011QSOVarSelection, PD2011SDSSstripe82HighZQSOVarSelection, Bruun2023VILLAIN, Nakoneczny2025QZO}. Quasars often exhibit significant stochastic variability, while stars show periodic variability over short timescales. However, the variability method requires long-term monitoring and well-performing facilities. If spectroscopy alone is insufficient to distinguish a DQ from an LQ, high-resolution deep imaging data are required to investigate whether the host galaxies of the two quasars exhibit a gas bridge or tidal tails connecting them \citep[e.g.,][]{LiuX2011LowZAGNPair, ChenYC2023CloseQPJ0749, Matsuoka2024MergingDQJ1215, ZhuYD2024QGpairALMAimaging, DengZJ2026LowZAGNPair, YueMH2026CloseQPJ2037}. Alternatively, dedicated photometric monitoring is required to obtain light curves of the two quasars and infer whether they display uncorrelated variability patterns.

The selection of LQ candidates is broadly similar to that of DQs. The difference is that, before obtaining spectra of the member sources in LQ candidates, color similarity is additionally required to remove candidates with large color differences between the members \citep[e.g.,][]{Oguri2006SQLS-Icolor, Agnello2017LQcolor, Agnello2019DESLQcolor, Dawes2023LScolorLQCBQC, HeZZ2023LScolorLQC, HeZZ2025KiDScolorCNNLQC, Chan2024SuGOHI-IX}. In addition, it is also necessary to check for the presence of a foreground massive deflector located between the two member quasars of an LQ candidate in the imaging survey data \citep[e.g.,][]{HeZZ2025KiDScolorCNNLQC}. After confirming that the member quasars have nearly identical spectra, one can examine the foreground deflector and its alignment with the two quasars. If the deflector is too faint or does not form a typical LQ-like configuration \citep[e.g.,][]{Bazzanini2025WSLQ18sepWHJ0400}, the system cannot be unambiguously confirmed as an LQ. In such cases, high-resolution deep imaging observations (e.g., HST and JWST) are also required to investigate whether the host galaxy of the lensed background quasar appears as a distorted arc induced by gravitational lensing \citep[e.g.,][]{Oguri2013WSLQJ1029+2623HST, Sharon2017WSLQJ2222+2745}. In some cases, it may also be necessary to perform mass modeling and source reconstruction for the system, thereby reproducing the observed data to confirm its LQ nature \citep[e.g.,][]{Andika2023LQCModeling, Vujeva2024SearchJWSTLensedhighZgalaxy, Barone2026GLCModeling}. One can also confirm the candidate system as an LQ by inferring reliable time delay(s) between the two quasar images, when high-quality light curves of the member quasars are available \citep[e.g.,][]{Haarsma1997LQB0957Timedelay, Pindor2005TimedelayLQconfrim, TA2023SpecImgTimedelayLQconfrim}.

In the era of large-scale spectroscopic surveys -- such as the current SDSS and surveys using DESI, LAMOST, and 4MOST \citep{deJong20194MOST}, as well as future JUST \citep{LiuCZ2024JUST}, MUST \citep{CaiZ2025MUST}, and CSST \citep{GongY2026CSST}, our spectroscopic follow-up faces both new challenges and opportunities. On the one hand, these spectroscopic surveys with fiber or slitless spectrographs acquire spectra at a rate far exceeding that of our follow-up efforts with slit spectrographs. On the other hand, the vast spectroscopic survey databases greatly accelerate the confirmation of DQs and LQs, allowing us to focus more on retrieving public spectra to address scientific topics. Meanwhile, a wealth of wide-field multi-band imaging surveys -- such as the current SDSS, Pan-STARRS, DESI-LS, and surveys using WFST \citep{WangTG2023WFST}, Euclid \citep{EuclidCollaboration2025EuclidOverview}, LSST \citep{Ivezic2019LSST}, and future CSST -- will not only provide sufficient multi-color data for quasar selection, but also offer multi-band images covering nearly the entire sky. This expands the search region of DQ and LQ candidates and enables spectroscopic follow-up for candidates that lie outside the footprints of those large-scale spectroscopic surveys.
%--------------------------------------------------------------------

\section{Summary} \label{sec5}
This work reported the spectroscopic follow-up of 11 QP candidates and the discovery of 6 DQs and 3 PQs. One of the DQs has a high redshift of $\sim$ 3.1. These candidates were selected from the MGQPC catalog proposed in the \textsc{Paper-I} and the early version of this catalog that was not published in \textsc{Paper-I}. Low-resolution long-slit spectroscopic observations were carried out employing 5 spectrographs equipped on 3 telescopes (OMR/XLT216, BFOSC/XLT216, YFOSC/LJ240, DBSP/P200, and NGPS/P200). The observed spectra were reduced using a Python module \texttt{PySpecRedux}, which was developed by us and is dedicated to low-resolution long-slit spectra. In addition, public spectra from SDSS DR17, DESI DR1, and LAMOST DR13 were also adopted as auxiliary data to improve the confirmation of the observed candidates. Two QSP systems were securely ruled out by our NGPS observations. By calculating $\lvert \Delta v_r \rvert$ between the two member quasars in each pair, 3 PQs were distinguished using a simple cut of $\lvert \Delta v_r \rvert$ $\leqslant$ 2000 km/s. The remaining 6 pairs with identical member redshifts were all confirmed as DQs through examination of the member quasar spectra, spectral flux ratios, and imaging data. The LQ hypothesis of 3 of the 6 DQs cannot be completely ruled out.

Three main topics were discussed to optimize the strategy for DQ confirmation and guide future spectroscopic follow-up and identification efforts. The first topic, concerning the color features of member B of the 11 observed targets, explained the reason why previous spectroscopic surveys missed several new quasars. The second topic, regarding photo-$z$ estimation, reported the photo-$z$ values for member B of five pairs. This acknowledged the positive role of the photo-$z$ estimation method in mitigating contamination from PQs and QSPs, while also highlighting the requirement for higher accuracy in reliable photo-$z$ estimation. The last topic reviewed and discussed the confirmation strategies for DQ and LQ candidates, and outlined future confirmation strategies for them in the context of the era dominated by large-scale spectroscopic and imaging surveys. These future confirmation strategies include retrieving public spectra from spectroscopic survey databases, performing spectroscopic follow-up for candidates located outside the footprints of large-scale spectroscopic surveys, acquiring high-resolution deep imaging data for candidates that already have spectra but remain ambiguous as DQs or LQs, obtaining high-quality light curves from photometric monitoring, and utilizing available public light curves from facilities such as WFST, LSST, and CSST. A recent work that confirmed 17 new DQs by retrieving public spectra from DESI DR1 will be reported in \textsc{Paper IV} of our series work.
%--------------------------------------------------------------------

\section*{Data Availability}
The confirmation outcomes (Table \ref{target_obs} and \ref{confirm_outcomes}) and spectroscopic data of the 6 DQs and 3 PQs are available at the CDS via \url{https://cdsarc.cds.unistra.fr/viz-bin/cat/J/A+A/}. Additionally, collaborative opportunities are open to conduct spectroscopic follow-up observations of the remaining MGQPCs.
%--------------------------------------------------------------------

% \section*{Acknowledgments}
\begin{acknowledgements}

% The authors thank the anonymous referee for the valuable comments that improved the quality and clarity of the manuscript. The authors thank Yiping Shu and Yuanzhen Han for their suggestions and guidance on several details. 
This work has been supported by the National Key R\&D Program of China (2021YFA0718500 and 2025YFA1614101) and by the Chinese National Natural Science Foundation grant No. 12333001. Z.H. acknowledges support from the China Postdoctoral Science Foundation under Grant Number GZC20232990 and the National Natural Science Foundation of China (Grant No. 12403104).

% XL216官方致谢
We acknowledge the support of the staff of the Xinglong 2.16m telescope. This work was partially supported by the Open Project Program of the Key Laboratory of Optical Astronomy, National Astronomical Observatories, Chinese Academy of Sciences.

% LJ240官方致谢
We also acknowledge the support of the staff of the Lijiang 2.4m telescope. Funding for the telescope has been provided by the Chinese Academy of Sciences and the People's Government of Yunnan Province.

% P200/TAP致谢
This research uses data obtained through the Telescope Access Program (TAP), which has been funded by the TAP association, including the Centre for Astronomical Mega-Science CAS (CAMS), XMU, PKU, THU, USTC, NJU, YNU, and SYSU.

% DESI-LS致谢
The DESI Legacy Imaging Surveys consist of three individual and complementary projects: the Dark Energy Camera Legacy Survey (DECaLS), the Beijing-Arizona Sky Survey (BASS), and the Mayall z-band Legacy Survey (MzLS). Pipeline processing and analyses of the data were supported by NOIRLab and the Lawrence Berkeley National Laboratory (LBNL). Legacy Surveys was supported by: the Director, Office of Science, Office of High Energy Physics of the U.S. Department of Energy; the National Energy Research Scientific Computing Center, a DOE Office of Science User Facility; the U.S. National Science Foundation, Division of Astronomical Sciences; the National Astronomical Observatories of China, the Chinese Academy of Sciences and the Chinese National Natural Science Foundation. LBNL is managed by the Regents of the University of California under contract to the U.S. Department of Energy.

% Pan-STARRS1致谢
The Pan-STARRS1 Surveys (PS1) and the PS1 public science archive have been made possible through contributions by the Institute for Astronomy, the University of Hawaii, the Pan-STARRS Project Office, the Max-Planck Society and its participating institutes, the Max Planck Institute for Astronomy, Heidelberg and the Max Planck Institute for Extraterrestrial Physics, Garching, The Johns Hopkins University, Durham University, the University of Edinburgh, and other institutions.

% SPARCL致谢
This research uses services or data provided by the SPectra Analysis and Retrievable Catalog Lab (SPARCL) and the Astro Data Lab, which are both part of the Community Science and Data Center (CSDC) program at NSF National Optical-Infrared Astronomy Research Laboratory. NOIRLab is operated by the Association of Universities for Research in Astronomy (AURA), Inc. under a cooperative agreement with the National Science Foundation.

% DESI DR1致谢
This research used data obtained with the Dark Energy Spectroscopic Instrument (DESI). DESI construction and operations are managed by the Lawrence Berkeley National Laboratory. This material is based upon work supported by the U.S. Department of Energy, Office of Science, Office of High-Energy Physics, under Contract No. DE–AC02–05CH11231, and by the National Energy Research Scientific Computing Center, a DOE Office of Science User Facility under the same contract. Any opinions, findings, and conclusions or recommendations expressed in this material are those of the author(s) and do not necessarily reflect the views of the U.S. National Science Foundation, the U.S. Department of Energy, or any of the listed funding agencies.

% SDSS DR17致谢
Funding for the Sloan Digital Sky Survey IV has been provided by the Alfred P. Sloan Foundation, the U.S. Department of Energy Office of Science, and the Participating Institutions. SDSS-IV acknowledges the support and resources from the Center for High Performance Computing at the University of Utah. The SDSS website is www.sdss4.org/. SDSS-IV is managed by the Astrophysical Research Consortium for the Participating Institutions of the SDSS Collaboration.

% LAMOST致谢
This work made use of the data from LAMOST (Large Sky Area Multi-Object Fiber Spectroscopic Telescope, also known as the Guoshoujing Telescope) (https://cstr.cn/31118.02.LAMOST). LAMOST is a Chinese national mega-science facility, operated by National Astronomical Observatories, Chinese Academy of Sciences.

We acknowledge the use of public imaging and spectroscopic data from the DESI Legacy Imaging Survey \citep{Dey2019DESILS}, Pan-STARRS \citep{Chambers2016PanSTARRS}, SDSS DR17 \citep{Abdurrouf2022SDSSDR17}, DESI DR1 \citep{DESICollaboration2025DESIDR1}, and LAMOST \citep{Cui2012LAMOST}. This research made use of the following software packages: Astropy \citep{AstropyCollaboration2013, AstropyCollaboration2018, AstropyCollaboration2022}, Astroscrappy\citep{McCully2018Astroscrappy, vanDokkum2001delCRLaplacianEdgeDetection}, Matplotlib \citep{Hunter2007Matplotlib}, NumPy \citep{Walt2011NumPy, Harris2020NumPy}, Pandas \citep{McKinney2010Pandas, pandas2022}, SciPy \citep{Jones2001SciPy, Virtanen2020SciPy}, HumVI \citep{Marshall2015SWHumVI, Marshall2016HumVI}, SPARCL \citep{Juneau2025SPARCL}.

\end{acknowledgements}

\bibliographystyle{aa}
\bibliography{aanda}
%--------------------------------------------------------------------

\begin{appendix}
\section{Quasa-star projections in this work} \label{appendixA}
\begin{figure}
    \centering
    \includegraphics[width=0.5\textwidth]{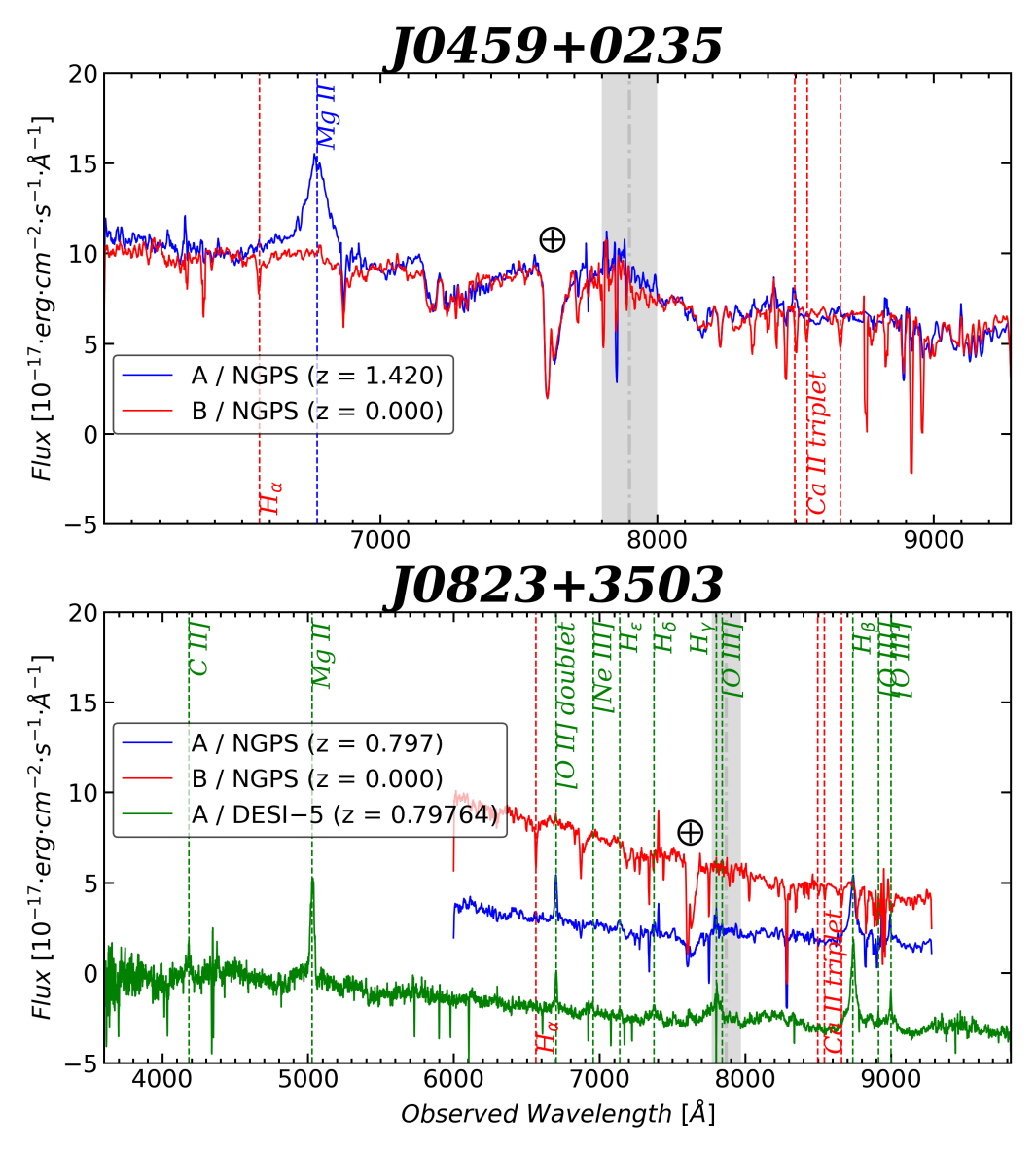}
    \caption{Similar to Figure \ref{DQ_Spectra} but for the two QSPs. The DESI DR1 spectrum of J0823+3503A observed on MJD 59586 is plotted, and its spectral flux is shifted by $-$5 units to avoid overlap. Both J0459+0235B and J0823+3503B exhibit absorption lines of H$\alpha$ and Ca II triplet, which are indicative of stellar features.}
    \label{QS_Spectra}
\end{figure}

The NGPS spectra revealed the stellar nature of J0459+0235B and J0823+3503B, that is, absorption lines of H$\alpha$ and Ca II triplet (Figure \ref{QS_Spectra}). Therefore, both pairs are QSP systems, despite their typical LQ-like configurations (see Figure \ref{MGQPC_Obs}).

\end{appendix}

\end{document}